\documentclass[a4paper,11pt]{article}
\usepackage{jcappub}

\pdfoutput=1

\usepackage{tabularx,booktabs}
\newcolumntype{Y}{>{\centering\arraybackslash}X}
\usepackage{fontawesome}
\usepackage[utf8]{inputenc}
\usepackage[font=small,labelfont=bf,tableposition=top]{caption}
\usepackage{float}
\usepackage{adjustbox}
\usepackage{arydshln}
\usepackage{empheq}

\usepackage{amsmath,amssymb,amsbsy,amstext, amsthm, simplewick, amsfonts, bm}

\usepackage[normalem]{ulem}
\usepackage{graphicx}
\usepackage{siunitx}
\usepackage{upgreek}
\usepackage{framed}
\usepackage{wrapfig}
\usepackage{multirow}
\usepackage{subfig}
\usepackage{bbm}
\usepackage[nameinlink, capitalise]{cleveref}
\usepackage{hyperref}
\usepackage[svgnames,dvipsnames,x11names]{xcolor}
\usepackage{wrapfig}
\usepackage{soul}
\usepackage[htt]{hyphenat}
\usepackage{orcidlink}

\usepackage{subcaption}

\title{Stochastic inflation and non-perturbative power spectrum beyond slow roll}

\author[1]{Devanshu Sharma\orcidlink{0009-0002-3302-2153}}

\affiliation[1]{Institute for Theoretical Particle Physics and Cosmology (TTK), RWTH Aachen University, \\ D-52056 Aachen, Germany}

\emailAdd{drsharma@physik.rwth-aachen.de}

\date{}

\abstract{Stochastic inflation, together with the $\Delta N$ formalism, provides a powerful tool for estimating the large-scale behaviour of primordial fluctuations. In this work, we develop a numerical code to capture the non-perturbative statistics of these fluctuations and validate it to obtain the exponential non-Gaussian tail of the curvature perturbations. We present a numerical algorithm to compute the non-perturbative curvature power spectrum and apply it to both slow-roll (SR) and ultra-slow-roll (USR) single-field models of inflation. We accurately generate a non-perturbative scale-invariant power spectrum in the SR scenario. In the USR case, we obtain a peak in the power spectrum that, in the time-independent regime, aligns with the structure of its perturbative counterpart. Additionally, We underscore how the evolving nature of the super-Hubble perturbations in the USR model complicates the numerical computation of the non-perturbative spectrum. }

\begin{document}

\begin{flushleft}
TTK-24-47
\end{flushleft}

\maketitle
\flushbottom
\section{Introduction}

Despite uncountable criticism and lack of smoking gun proof, inflation stands as arguably the most established theory explaining the creation of primordial perturbations that gave rise to inhomogeneous structures in the present universe, even decades after its discovery \cite{Starobinsky:1980te,Guth:1980zm,Albrecht:1982wi, Guth:1980zm, Linde:1981mu}. A number of predictions have been successfully verified by either ground-based or space-based missions viz. the flat geometry of the spacetime at very large scales \cite{Planck:2015fie, Planck:2018vyg}, statistical homogeneity, isotropy and scale invariant adiabatic primordial density perturbations from observing the spatial temperature anisotropies of the Cosmic Microwave Background (CMB) \cite{Planck:2013jfk, Planck:2015sxf,Planck:2018jri}, with forthcoming measurements in line to detect and measure the primordial gravitational waves from B-modes of the CMB \cite{Matsumura:2013aja, Hazumi:2019lys, LiteBIRD:2022cnt, CMB-S4:2016ple, Abazajian:2019eic}.

Several inflationary models with a slow-roll (SR) attractor solution have been devised that give rise to scale-invariant Gaussian perturbations \cite{Lucchin:1984yf, Sahni:1990tx, Cao:2002qe, Boubekeur:2005zm}. On the other hand, a non-attractor regime in the inflaton potential can be featured by introducing a flat region in the potential, leading to an Ultra-Slow roll (USR) regime, which enhances the perturbations by many orders of magnitude at the scales yet unprobed by the present measurements. An important consequence of this is that such large fluctuations could undergo gravitational collapse after re-entering the horizon post-inflation resulting in Primordial Black Holes (PBHs) \cite{Zeldovich:1967lct, Hawking:1971ei, Chapline:1975ojl, Carr:1974nx, Garcia-Bellido:2017mdw, Ballesteros:2020qam, Gangopadhyay:2021kmf, Mishra:2019pzq, Stamou:2024lqf}. Since the detection of gravitational waves from binary black hole mergers by the LIGO-VIRGO collaboration \cite{LIGOScientific:2016aoc, LIGOScientific:2016dsl}, many eyes have turned towards PBHs as potential candidates. The possibility that PBHs could contribute to a notable fraction of present dark matter density has still not been ruled out, given that there are still some unconstrained mass windows \cite{Bird:2016dcv,Carr:2023tpt, Carr:2020gox, Escriva:2022duf, Riotto:2023gpm}. Moreover, such large scalar perturbations couple to tensor perturbations at second order in perturbation theory, resulting in the generation of Scalar Induced Gravitational Waves that might serve as a viable explanation for the recent detection of stochastic gravitational wave background by Pulsar Timing Array \cite{EPTA:2023fyk, EPTA:2023sfo,EPTA:2023xxk} from several collaborations of NANOGrav worldwide \cite{NANOGrav:2023gor, NANOGrav:2023hde, Ellis:2023oxs}.

It is known that the large amplitude fluctuations resulting from USR models may not follow strict Gaussian statistics. The assumption of Gaussianity could have serious implications in computing the abundance of PBHs, which form from rare large fluctuations and hence, are susceptible to the tail of the probability distribution of the curvature perturbations \cite{Taoso:2021uvl, Gow:2022jfb,Pi:2024jwt, Pattison:2017mbe, Ezquiaga:2019ftu, Ferrante:2022mui}. Stochastic inflation was introduced to incorporate the non-linear effects resulting from the influence of the quantized small-scale modes of the field fluctuations on the large-scale dynamics of the expanding background, also called quantum diffusion \cite{Starobinsky:1994bd, Habib:1992ci, Salopek:1990jq, Mijic:1994vv, Tsamis:2005hd}. The stochastic approach treats the overall volume in Fourier space as large-scale modes that have classicalized substantially and small-scale modes that influence the large-scale modes every time they cross the threshold.

Two tools are generally utilized to study the non-perturbative statistics of the inflationary perturbations viz. the separate universe approach and the $\Delta N$ formalism. The separate universe approach treats every super-Hubble patch as a local FLRW universe. On the other hand, the $\Delta N$ formalism uses the separate universe approach and relates the differences in expansion rates of such independent FLRW universes to the total comoving curvature perturbation, leading us to the perturbations without solving the linearized perturbed Einstein equations  \cite{Sasaki:1986hm,Mukhanov:1988jd}. When the $\Delta N$ formalism is applied after infusing the quantum diffusion effects onto the classical background evolution, the method is called stochastic $\Delta N$ formalism.

Lately, numerous works have studied non-perturbative statistics via stochastic inflation for both SR and USR models of inflation. It was shown in \cite{Figueroa:2020jkf, Figueroa:2021zah, Ezquiaga:2019ftu, Pattison:2021oen, Animali:2022otk, Tomberg:2023kli,Pi:2022ysn} that the tail of the probability distribution of the curvature perturbation exponential non-Gaussianity, thus predicting a much higher PBH abundance when compared with the Gaussian counterpart. Moreover, \cite{Jackson:2022unc, Jackson:2024aoo, Tomberg:2022mkt} applied importance sampling technique to significantly speed up the computation of the tail of the distribution by cherry-picking more rare realizations. On the other hand, a non-perturbative framework of computing the curvature power spectrum has also been coined \cite{Fujita:2013cna, Fujita:2014tja, Ando:2020fjm} and also for higher correlators \cite{Vennin:2015hra}. It has been exhibited that quantum diffusion effects on the power spectrum of USR models lead to significant deviations from the linearized treatment in the beyond SR models \cite{Biagetti:2018pjj, Ezquiaga:2018gbw, Cruces:2018cvq, De:2020hdo, Cruces:2021iwq}. Moreover, Ref. \cite{Animali:2024jiz} has studied the effects of non-Gaussianity arising from the quantum diffusion on the clustering of PBHs and also explained the significance of volume weighting when computing the two-point statistics of the curvature perturbations.

At this stage, it is important to highlight two specific works due to their proximity to our results. First, Ref. \cite{Biagetti:2018pjj} analyzes quantum diffusion effects during a non-attractor phase beyond slow-roll for several realistic inflationary models, and shows the impact of stochastic noise from sub-Hubble modes, computed using an approximation, on the curvature power spectrum at the end of inflation. For this purpose, they compute the Mukhanov-Sasaki equation numerically over a quantum-corrected background. Moreover, they set up a statistical distribution of the field velocity by running many realizations of stochastic inflation using the Kramers-Moyal equation, a generalized form of the Fokker-Planck equation, again using the simplistic approximation of the noise variance. On the other hand, Ref. \cite{Ezquiaga:2018gbw} computes the correlators of the field perturbations via the time derivative of the solution to the Fokker-Planck equation and numerically solves for the system of equations. The power spectrum is obtained as a time derivative of the two-point correlators without simulating many stochastic realizations. Our work is different from both of them in regards that we treat our model via the Langevin equations of linear order, and numerically simulate the inflationary regime taking into account the stochastic background evolution. We repeat this for several realizations to catch the first passage time of the field through its value at the end of inflation in each of them. Finally, we make use of the $\Delta N$ formalism to first study the probability distribution of the curvature perturbations and then the non-perturbative power spectrum using the algorithm explained in \cref{non_pert_algo}. To the best of our knowledge, such a numerical implementation of the non-perturbative scalar power spectrum, accounting for the numerical evolution of the stochastic noise, has not been done in the context of USR models.

This paper is drafted as follows: In \cref{Perturbative_intro} we summarize the picture of a homogeneous FLRW universe with the first-order perturbations over it. In \cref{Stochastic_delta_N} we give a formal intuition of stochastic inflation, $\Delta N$ formalism and collect the important equations. \cref{testing_section} is reserved for testing our numerical code to obtain the probability distribution of perturbations and check it against the known results. Our main results are presented in \cref{non_pert_algo}, where we elucidate the numerical approach to non-perturbatively getting the power spectrum, accompanied by its application on attractor and non-attractor models. We end this discussion with concluding remarks in \cref{conclusion}. In \cref{SW_inaccuracy}, we recall the computation of the perturbed Klein-Gordon equation to point out an inaccuracy in the equation of motion for sub-Hubble perturbation evolution in Ref. \cite{Figueroa:2020jkf, Figueroa:2021zah}. For the more curious readers, we provide the visuals of the stochastic noise and the accuracy check of our code in \cref{noise_evolution_plots} and \cref{accuracy check_appendix} respectively.

\section{Linearized Perturbations}\label{Perturbative_intro}

%

\noindent Assuming a flat Friedmann–Lemaître–Robertson–Walker (FLRW) metric, the background expansion rate during inflation and the Klein-Gordon equation for the field evolution are respectively,

\begin{align}
    H^2 (t) &=V(\phi) + \frac{1}{2}\partial_{t}\phi \ \partial^{t}\phi~, \label{Friedmann_eq}\\
    \Ddot{\phi}(t) &= - 3 H \dot{\phi}(t) - \frac{\partial V}{\partial \phi (t)}~, \label{KG_eq}
\end{align}

\noindent where $V(\phi)$ is the potential energy of the field and the subscript `$t$' denotes the derivative with respect to coordinate time. As there is no other degree of freedom during single-field inflation, \cref{Friedmann_eq,KG_eq} together with suitable initial conditions are sufficient to study the unperturbed dynamics of the universe. In the presence of small perturbations, the homogeneous and perturbed parts of the field can be treated separately, 

\begin{align}\label{pert_inflaton}
    \phi ({\rm x},t) \approx \bar{\phi} (t) + \delta \phi ({\rm x},t) ~.
\end{align}

\noindent The field perturbations follow the perturbed Klein-Gordon equation (See \cref{SW_inaccuracy} for details) which, after being taken to the Fourier space become

\begin{align}\label{pert_KG_eq}
\delta \phi_{\rm \textbf{k}}^{\prime \prime} +  (3 + H^{\prime}/H) \delta \phi^{\prime}_{\rm \textbf{k}} + \Theta^2 \delta \phi_{\rm \textbf{k}} &=0~,
\end{align}

\noindent with the oscillation frequency $\Theta^2 = {\rm k}^2 /(a H)^2+\bar{\phi}_{\rm k}^{\prime^2} \left(3+H^{\prime} / H\right)+2 \phi^{\prime}_{\rm k} V_{, \bar{\phi}} / H^2+V_{, \bar{\phi} \bar{\phi}} / H^2$ corresponding to the wavenumber ${\rm k} = \left|{\rm \textbf{k}}\right|$. The Bunch-Davies vacuum conditions are generally the standard choice for the initial conditions. A convenient way to characterize perturbations is via a gauge invariant variable, called comoving curvature perturbation $\mathcal{R}_{\rm k}$, whose two-point statistics is expressed in terms of the dimensionless power spectrum $\mathcal{P_R}({\rm k})$ defined as
\begin{align}
    \langle \mathcal{R}_{\rm \textbf{k}}  \mathcal{R}_{\textbf{k}^{\prime}} \rangle &= \frac{{\rm k}^3}{2 \pi^2} \mathcal{P_R}({\rm k}) \delta (\textbf{k} - \textbf{k}^{\prime})
\end{align}

\section{Stochastic $\Delta$N Formalism}\label{Stochastic_delta_N}

\subsection{A quick guide to Stochastic Inflation}\label{Stochastic_intro}

With the aim of incorporating the sub-Hubble quantum effects during inflation to the homogeneous background at super horizon scales, stochastic inflation involves splitting the field and its velocity into a short wavelength (SW) field and a long wavelength (LW) field such that
\begin{align}
     \left[\phi (\rm{x},N),  \pi (\rm{x},N)\right] &= \left. \left[ \bar{\phi}(N),  \bar{\pi}(N)\right] \right|_{\rm LW} + \left. \left[ \delta \phi(\rm{x}, N), \delta \pi(\rm{x}, N) \right] \right|_{\rm SW}~, \\
    \left. \left[ \delta \phi(\rm{x}, N), \delta \pi(\rm{x}, N) \right] \right|_{\rm SW}  &= \frac{1}{2 \pi^{3/2}} \int _{{\rm k}>k_{\rm cg}}  \left[ \phi_{\rm \textbf{k}} (N), \pi_{\rm \textbf{k}} (N) \right]  e^{-i \textbf{k} \cdot x} d^3 {\rm k}~, \\
    \left. \left[ \bar{\phi}(N),  \bar{\pi}(N)\right] \right|_{\rm LW}  &= \frac{1}{2 \pi^{3/2}} \int_ {{\rm k}<k_{\rm cg}} \left[ \phi_{\rm \textbf{k}} (N), \pi_{\rm \textbf{k}} (N) \right] e^{-i \textbf{k} \cdot x} d^3 {\rm k} ~,
\end{align}
where $\bar{\pi} = \frac{d \bar{\phi}}{d \bar{N}}$  is the inflaton velocity and, ${\rm k}_{\rm cg} = \sigma a H$  is the coarse-graining scale that separates the LW and SW modes. The former is a classical variable that evolves with the classical equations of motion, whereas, the latter retains its quantum properties and hence has a stochastic nature obeying \cref{pert_KG_eq}. The parameter $\sigma$ thus determines the separation scale for classicalized background $({\rm k} \ll {\rm k}_{\rm cg})$ from the quantum perturbations $({\rm k} \gg {\rm k}_{\rm cg})$. The choice of $\sigma$ is not significant for the SR model as the perturbations essentially freeze after horizon exit. However, this is not the case with USR models where the perturbations evolve even at the superhorizon scales \cite{Figueroa:2021zah, Casini:1998wr, Matarrese:2003ye}. We refrain from going into the details of this issue and choose a considerably small value of $\sigma$ (or a separation scale considerably larger than the physical horizon) such that the modes can be treated as classical, but not too small otherwise, we will not estimate the small scale effects on the background correctly (See also \cite{Figueroa:2020jkf, Polarski:1995jg, Grain:2019vnq}). Keeping this in mind, we set $\sigma=0.01$ throughout this text. We shall drop the subscripts LW/SW on the variables from now on unless otherwise stated. After one substitutes the coarse-grained field and velocity in the Klein-Gordon equation \cref{KG_eq} \cite{Pattison:2019hef, Figueroa:2020jkf} one obtains the following evolution for the background
\begin{align}
    {\bar{\phi}}^{\prime} &= \bar{\pi}+ \xi_{\phi} \label{phi_eq}\\
    {\bar{\pi} }^{\prime} &=  - (3 - \epsilon) \bar{\pi} - V_{, \bar{\phi}}/ H^2 +  \xi_{\pi}~, \label{pi_eq}
\end{align}
with $\epsilon = \frac{-H^{\prime}}{H}$ being the first slow-roll parameter. The quantities labelled with $\xi$ are field and momentum noise terms 
\begin{align}\label{noise_eq}
    \xi_{\phi} (N) &=  \frac{1}{2 \pi^{3/2}} \int \beta^{\prime}({\rm k}- \sigma aH)  \phi_{\rm k} (N)  e^{-i {\rm \textbf{k}} \cdot x} d^3 {\rm k}~, \\\ 
    \xi_{\pi} (N) &=  \frac{1}{2 \pi^{3/2}} \int \beta^{\prime}({\rm k}- \sigma aH)  \pi_{\rm k} (N)  e^{-i {\rm \textbf{k}} \cdot x} d^3 {\rm k}~.
\end{align}
For the window function $\beta({\rm k})$, we pick the traditionally used Heaviside step function that introduces a sharp cutoff at the coarse-graining scale \cite{Starobinsky:1994bd}. It is worth noting that the choice of window function has important implications particularly when the potential has a non-attractor regime (Refer \cite{Winitzki:1999ve, Casini:1998wr, Grain:2017dqa, PhysRevD.42.1027} for further discussion). The SW contribution coming as a stochastic noise term modifies the background evolution each time a SW mode crosses ${\rm k}_{\rm cg}$. Assuming SW perturbations to be vacuum fluctuations, the noise has a Gaussian distribution centred at zero i.e. $\langle \xi_{\phi} (N) \rangle = \langle \xi_{\pi} (N) \rangle = 0$. The two-point noise correlators are to be derived by promoting the field and its velocity as quantum operators in the ultraviolet limit, $\left[ \delta \hat{\phi}_{\rm \textbf{k}}(\rm{x}, N), \delta \hat{\pi}_{\rm \textbf{k}}(N) \right] = \hat{a}_{\rm \textbf{k}}  \left[\phi_{\rm k} (N),  \pi_{\rm k} (N)\right]  + \hat{a}_{\rm \textbf{k}}^{\dagger}  \left[\phi_{\rm k} (N),  \pi_{\rm k} (N)\right]$ from which results the following  \cite{Pattison:2021oen, Mishra:2023lhe}
\begin{align}
    \langle 0 \left| \hat{\xi}_{\phi} (N_{\rm x}) \hat{\xi}_{\phi} (N_{\rm y}) \right| 0 \rangle &= \frac{d(\sigma a H)^3}{6 \pi^2 d N} \left. \left| \delta \phi_{\rm \textbf{k}} \right|^2 \right |_{{\rm k}=\sigma a H} \delta (N_{\rm x} - N_{\rm y}) \label{field_noise_variance_SW}\\
    \langle 0 \left|  \hat{\xi}_{\pi} (N_{\rm x}) \hat{\xi}_{\pi} (N_{\rm y}) \right| 0  \rangle &=  \frac{d(\sigma a H)^3}{6 \pi^2 d N} \left. \left| \delta \phi_{\rm \textbf{k}}^{\prime} \right|^2 \right |_{{\rm k}=\sigma a H} \delta (N_{\rm x} - N_{\rm y})~,\label{momentum_noise_variance_SW}       \\
    \langle 0 \left| \hat{\xi}_{\phi} (N_{\rm x}) \hat{\xi}_{\pi} (N_{\rm y}) \right| 0  \rangle &=    \frac{d(\sigma a H)^3}{6 \pi^2 d N} \left. \delta \phi_{\rm \textbf{k}} \delta \phi_{\rm \textbf{k}}^{\prime}  \right |_{{\rm k}=\sigma a H} \delta (N_{\rm x} - N_{\rm y})~, \label{cross_noise_variance_SW}      \\
    \langle \delta \phi^2 \rangle &= {\rm Re} \left(\frac{\delta \phi_{\rm \textbf{k}}}{\delta \phi^{\prime}_{\rm \textbf{k}}} \right) \langle \delta \pi^2 \rangle ~,\label{squeeze_field_momentum_SW}
\end{align}
Some important notes: firstly, the noise is computed at the instant when the mode crosses ${\rm k}_{\rm cg}$, which is sufficiently larger than the physical horizon from our choice of $\sigma$. The SW Fourier modes are treated as quantum harmonic oscillators, each corresponding to the wavelength $\frac{2 \pi}{{\rm k}}$, whose evolution is tracked by the Mukhanov-Sasaki equation. But when allowed to evolve for sufficient time post horizon exit, we make sure that the position and momentum uncertainty of the fluctuations is negligible and hence, at ${\rm k}={\rm k}_{\rm cg}$, the modes are eligible to receive classical treatment (notice that there are no hats on the noises in \cref{phi_eq,pi_eq}). Another consequence of the classicalization of the SW modes is that the field and momentum noise become highly correlated due to squeezing (for more on squeezing and classicalization, refer \cite{Burgess:2006jn, Martin:2012ua, Grishchuk:1990bj, Polarski:1995jg, Martineau:2006ki}). The delta function in the \cref{field_noise_variance_SW,momentum_noise_variance_SW,cross_noise_variance_SW} comes from the derivative of the Heaviside window function, hence, selecting a steplike separation scale ensures an instantaneous impact of the noise on the background. We precompute the noise correlators on a classical background and use them to evolve the stochastic Langevin equations for several realizations. In principle, the fully non-linear approach involves incorporating the backreaction of the SW evolution on the correlators, thus treating the noise as non-Markovian. This is computationally expensive. We leave the comparison of our results with the full approach for future work. Computing the noise correlators on a classical background is not an unreasonable supposition as it has been shown that the backreaction effects show a little deviation from the Markovian approach while handling the statistics of perturbations via the stochastic formalism   \cite{Jackson:2024aoo, Ballesteros:2017fsr, Figueroa:2020jkf}.

\subsection{$\Delta N$ formalism }
In this section, we offer a passing mathematical base of the $\Delta N$ formalism, which is one of the most common non-perturbative approaches to studying the primordial perturbations (refer \cite{Starobinsky:1986fx, Wands:2000dp,Vennin:2024yzl} for more details). The perturbed line element for a generalized Friedman-Robertson-Walker (FRW) metric in a vanishing curvature is given by \cite{Mukhanov:1990me, Kodama:1984ziu, Bardeen:1980kt, Lifshitz:1945du}


\begin{align}\label{pert_FRW_metric}
    ds^2 &= (1+ 2A) dt^2 + 2 a^2(t) \nabla_{\mu} B dx^{\mu} dt + a^2(t) (\left[ 1 - 2 \psi \right] \gamma_{\mu \nu} + 2 \nabla_{\mu} \nabla_{\nu} E) dx^{\mu} dx^{\nu}~,
\end{align}

%
%
\noindent the nabla operator stands for a covariant derivative and $\gamma_{\mu \nu}$ is the spatial metric at zeroth order. A contravariant four-vector field can be introduced that is purely timelike to the metric \cref{pert_FRW_metric}

\begin{align}
    v^x &= (1- A, - \nabla^{\mu} B)~,
\end{align}

\noindent and orthogonal to the spacelike hypersurfaces on which we wish to find the nature of the curvature perturbation. One can take the covariant derivative of the timelike vector field $v^x$ with respect to the coordinate time to obtain the perturbed expansion rate $\Theta$ of the spacelike hypersurfaces


\begin{align}\label{expansion_rate_geometric}
    \Theta &= \nabla_{\rm x} v^x = 3 H  - 3 \dot{\psi} + \nabla^2 (\dot{E} - B)~,
\end{align}

\noindent after making the metric perturbation as subject and subtracting the background expansion from the total expansion rate i.e. $\Delta \Theta = \Theta - 3 H$, we get

\begin{align}
    \dot{\psi} &= - \frac{\Delta \Theta + \nabla^2 (\dot{E} - B)}{3}~.
\end{align}


\noindent Now consider a spatial hypersurface $\Lambda (t_1)$ at some initial time $t_1$ in the uniform expansion gauge $(\delta \rho = 0)$ containing two large patches M and N, having sizes $\lambda_M$ and $\lambda_N$ respectively, separated by a physical distance $\lambda$ ($\gg \lambda_M , \lambda_N$). These two patches can be regarded as independent homogeneous FLRW spacetimes with scales larger than the physical horizon. Given that the metric perturbation, corresponding to the scale $\lambda$ on this hypersurface is $\Delta \psi_1 = \psi_{M_1} - \psi_{N_1}$, the metric perturbation at some future spatial hypersurface $\Lambda (t_2)$ becomes

\begin{align}
    \Delta \psi_2 &= \Delta \psi_1 - \Delta N~,
\end{align}

\noindent with $\Delta N$ being the difference in the expansion of M and N. The metric perturbation coincides with the comoving curvature perturbation $\mathcal{R}$ in uniform expansion gauge, which for an adiabatic perturbation equals setting the field perturbation to zero. Hence,




\begin{align} \label{eq:R_N_relation}
   \Delta \mathcal{R}&= \Delta N~,
\end{align}

\noindent where $\Delta \mathcal{R} = (\Delta \psi_{M} - \Delta \psi_{N})$. This is the essence of the $\Delta N$ formalism, where the differential expansion rate across various locally FLRW universes is equal to the comoving curvature perturbation corresponding to the separation scales. Recently, it was noted that the $\Delta N$ formalism suffers from certain limitations when the inflaton potential features a non-attractor phase. For instance, Ref. \cite{Jackson:2023obv} shows that when entering or exiting the USR regime, the spatial gradients do not vanish at some finite super-Hubble scales because of the presence of entropy perturbations \cite{Leach:2001zf}. The full statistical behaviour of the curvature perturbations in this regime can then be obtained by adding the matched corrections to the classical $\Delta N$ method. Another argument has been made by Ref. \cite{Cruces:2018cvq} that during the USR phase, the stochastic equation of motion for the coarse-grained infrared part of the inflaton at large scales only coincides with the equation of motion of perturbations from quantum field theory at linear order in the slow roll parameters, whereas, if one includes the momentum constraint in the stochastic formalism, both cases agree up to all orders in SR parameters \cite{Cruces:2022imf,Cruces:2024pni}. More recently, Ref. \cite{Artigas:2024xhc} has discussed that it is possible to correctly account for the curvature perturbations during the ${\rm SR} \Leftrightarrow {\rm USR}$ transition if one considers the gradient corrections to the classical $\Delta N$ formalism and in this case, the coarse-graining scale does not have to be much larger than the physical horizon. For the purposes of this work, we will assume the validity of the $\Delta N$ formalism without such corrections throughout the inflationary phase and employ it alongside stochastic inflation to compute the two-point statistics of perturbations at super-Hubble scales.


\section{Probability distribution of the first passage time}\label{testing_section}
\subsection{Toy potential}
We begin our analysis by simulating several realizations of a toy model, first studied in \cite{Ezquiaga:2019ftu, Figueroa:2021zah}, to generate the probability distribution function (PDF) of the first passage time of the end of inflation. The tilted-shaped potential is given as
\begin{align}\label{toy_potential}
    V (\phi) &= V_0 \left( 1 + \alpha (\phi - \phi_0) + \beta (\phi - \phi_0)^3\right)~,
\end{align}

\noindent with the parameter values $\left[ V_{0}, {\beta}, {\alpha},  \phi_{0}  \right]= \left[\frac{3}{25}\pi^2, 10^{-3},  \{2,4\} \times 10^{-2}, \frac{10^{-1}}{\alpha} \right]~.$


\noindent In the simulation we precompute the background evolution using the system of equations \cref{Friedmann_eq,phi_eq,pi_eq}. Then we evolve the perturbations over this background and compute the variance of the stochastic kicks for the Fourier modes when they cross the coarse-graining scale. Finally, we run the stochastic realizations of the background evolution using the stochastic Runge-Kutta method of order two keeping a constant step size. The field starts rolling down from the flat section of the potential, $\phi_{0}$, till it reaches zero. First, to reproduce the results of \cite{Ezquiaga:2019ftu, Figueroa:2021zah}, we employ a so-called simplified treatment of the noise variance, where a saturated super-Hubble magnitude of SW modes goes into the noise correlator and the momentum noise is ignored such that after substituting $\left. \left| \delta \phi_{\rm \textbf{k}} \right|^2 \right |_{{\rm k} \gg a H} = \frac{H^2}{2 {\rm k}^3}$,  \cref{field_noise_variance_SW,momentum_noise_variance_SW,cross_noise_variance_SW} become

\begin{align}
    \langle 0 \left| \hat{\xi}_{\phi} (N_{\rm x}) \hat{\xi}_{\phi} (N_{\rm y}) \right| 0 \rangle &= \frac{H^2}{4 \pi^2} ~,\ \delta (N_{\rm x} - N_{\rm y}) ~,\label{simplified_field_noise_variance_SW}\\
    \langle 0 \left| ~, \hat{\xi}_{\pi} (N_{\rm x}) \hat{\xi}_{\pi} (N_{\rm y}) \right| 0  \rangle &=  0 ~,\label{simplified_momentum_noise_variance_SW}       \\
    \langle 0 \left| \hat{\xi}_{\phi} (N_{\rm x}) \hat{\xi}_{\pi} (N_{\rm y}) \right| 0  \rangle &=    0 ~.\label{simplified_cross_noise_variance_SW}
\end{align}

\begin{figure}[htb]
    \centering
    \includegraphics[width=\linewidth]{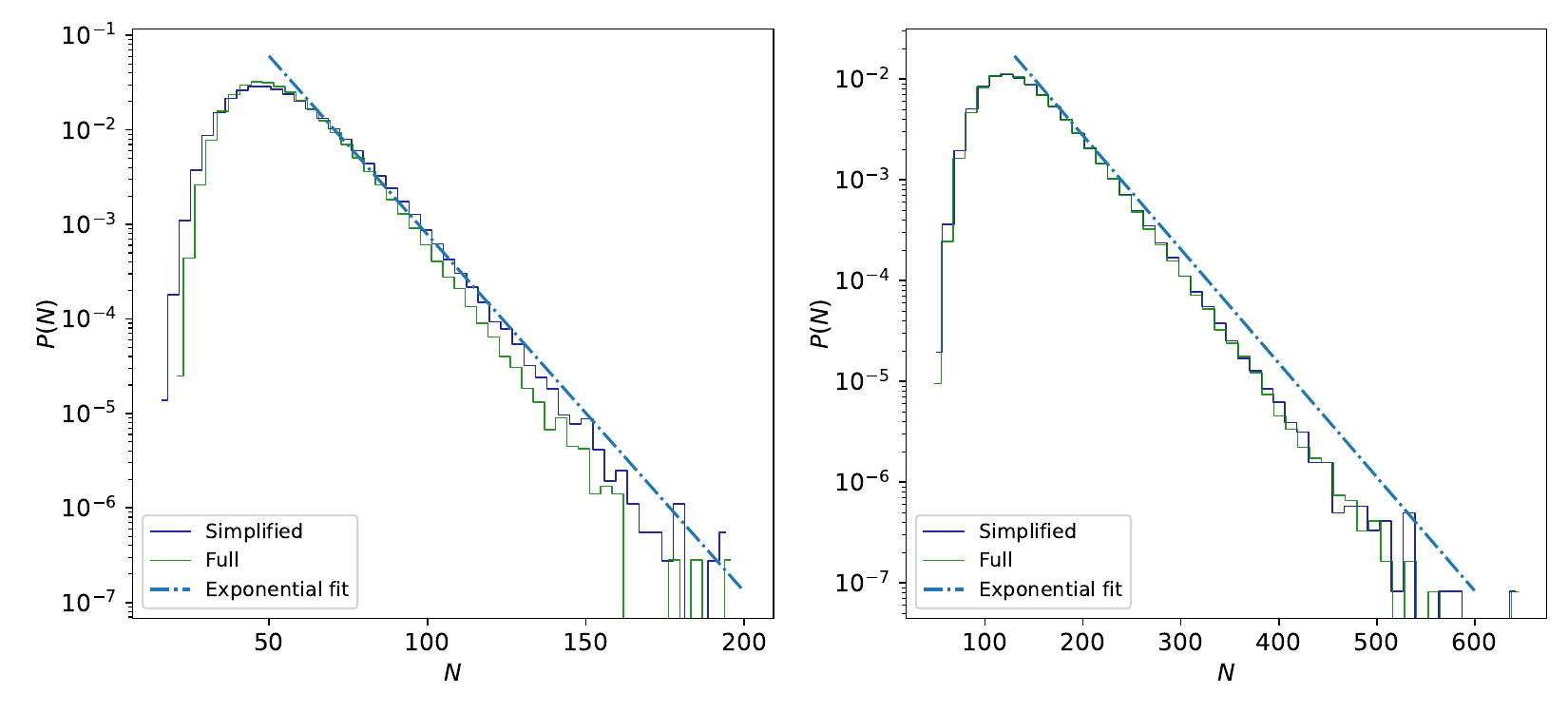}
    \caption{\textbf{Left Panel:} Probability distribution function of the end of inflation e-foldings for the potential \cref{toy_potential}, computed using full calculation (green curve) and compared with the simplified case (dark blue curve), for $\alpha=0.04$. Exponential fit from \cite{Figueroa:2021zah} is also shown as the dotted-dashed line. \textbf{Right Panel:} Same as in the left panel but for $\alpha=0.02$.}
    \label{fig:PDF_toy_model}
\end{figure}

\noindent Then, we treat the same models with full noise exposure. Both the results are depicted in \cref{fig:PDF_toy_model}. The exponential fit of the type $f(a)\propto  e^{-D a}$ by the dotted-dashed line proves the presence of a non-Gaussian tail in the probability distribution of $N$. Upon normalizing the distribution, we find that the simplified method gives the respective slopes for $D \left(\alpha=0.04, \alpha= 0.02 \right) \approx \left[0.086, 0.027 \right] $. The first case agrees well with \cite{Ezquiaga:2018gbw,Figueroa:2021zah} up to a few per cent accuracy. In the second case viz. $\alpha=0.02$, our distribution predicts a smaller variance of $N$ and this feature is insensitive to precision settings. We attribute this to the backreaction of the noise that enhances the stochasticity of the system leading to a wider distribution. This effect is negligible in the $\alpha=0.04$ case due to the shorter duration of inflation but, its cumulative behaviour has become more noticeable when a broad range of modes cross the coarse-graining scale and contribute as noise.
\noindent On the other hand, the full method starts kicking the background only after the largest mode in the simulation crosses the coarse-graining scale and since the flat section is where we start from, we miss out on the noise effects here. This is why, for the case $\alpha=0.04$, this method has a slightly lower variance than the simplified method. Interestingly, in the other case, the full treatment complies with the simplified results. This is again due to the extended duration of inflation, because of which the effects of kicks not being added in the first few e-folds get nullified. We also confirm the suppressed behaviour of the momentum noise and thus, the good agreement of the simplified method for the toy models (See \cref{noise_evolution_plots}). Overall, we simulate $4 \times 10^6$  realizations for this model.



\subsection{Realistic potential: USR feature}\label{realistic_potential_section}
As another independent check for our code, we simulate the model analysed in \cite{Ballesteros:2024pwn, Ballesteros:2020qam} which involves a polynomial potential and an inflaton field non-minimally coupled to gravity,

\begin{align}\label{eq:realistic_potential}
    V(\phi) &= \frac{V_0 \phi^4}{4(1 + A_1 \phi^2)^2} \left[ 3 + A_1^2 \phi_0^4 - 8(1 + A_2) \frac{\phi_0}{\phi} + 2(1 + A_3)(3 + A_1 \phi_0^2) \frac{\phi_0^2}{\phi^2} \right]~,
\end{align}

\noindent where the field $\phi$ has a non-canonical kinetic term in the Einstein frame. To work with a canonically normalized field $\Phi$, one can perform the following change of variable

\begin{align}
    \Phi &= \sqrt{\frac{1 + 6A_1}{A_1}} \sinh^{-1} \left[ \phi \sqrt{A_1(1 + 6A_1)} \right] - \sqrt{6} \tanh^{-1} \left[ \frac{\sqrt{6} A_1 \phi}{\sqrt{1 + A_1(1 + 6A_1)\phi^2}} \right]~,
\end{align}

\noindent We set the parameters to be the same as in \cite{Ballesteros:2024pwn} i.e. $A_2 = 0.0089$, $A_3=0.011$, $A_1=0.325479$ which adjust the inflection point at $\phi=1$. We normalize $V_0$ to get the correct curvature power spectrum amplitude on Planck scales.
\begin{figure}[htb]
    \centering
    \includegraphics[width=\linewidth]{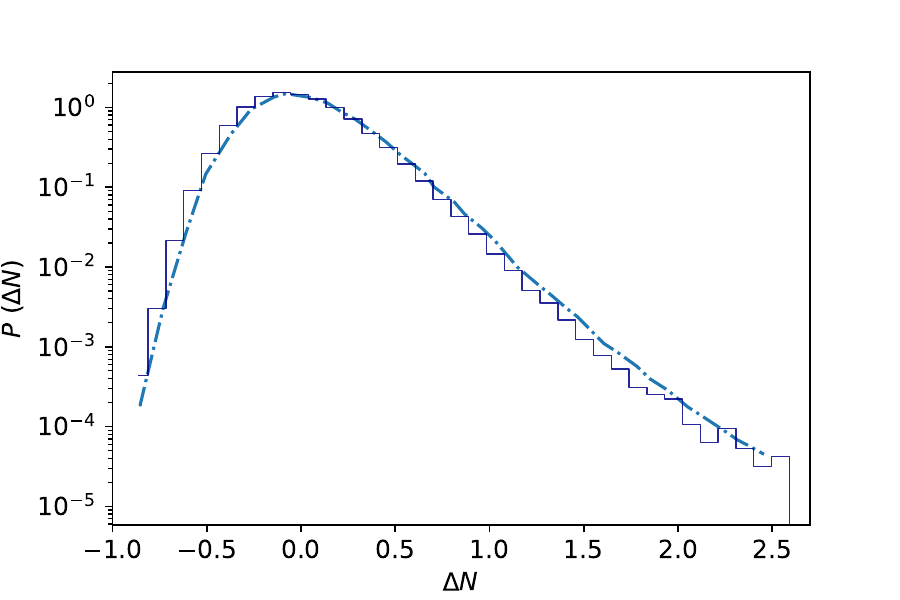}
    \caption{Probability distribution $\Delta N = N - \bar{N}$ for the potential with an USR regime \cref{eq:realistic_potential}. The exponential fit (dotted dashed line)  is taken from Figure 5 of \cite{Ballesteros:2024pwn}. The chosen parameter set is $\left[A_1, A_2, A_3, \phi_0,  \right] = \left[0.325479, 0.0089, 0.011, 1 \right]$ and $V_0$ normalizes the CMB observations. }
    \label{fig:PDF_realistic}
\end{figure}
\noindent The introduction of the non-attractor regime demands more extensive computations in order to resolve the field dynamics in the flat section of the potential. The result after running $\sim 10^6$ stochastic realizations is shown in the \cref{fig:PDF_realistic}. We were able to reach up to $\Delta N=2.5$ deep inside the tail of the distribution. We confirm the agreement of our result with the exponential tail in Figure 4 of \cite{Ballesteros:2024pwn} up to a few per cent accuracy.



\section{Algorithm for the non-perturbative power spectrum}\label{non_pert_algo}

\noindent It is known that the perturbations have to be of the order $\mathcal{R} \sim \mathcal{O}(10^{-1})$ to generate a significant abundance of PBHs, at very small scales compared to CMB scales. In the presence of such large perturbations, the scales do not remain independent and hence lose their Gaussian characteristic as well. The perturbative treatment \cref{eq:mukhanov_sasaki} is linear, does not account for such interactions and thus misses the non-Gaussian nature of the large fluctuations. On the other hand, the non-perturbative treatment via the $\Delta N$ formalism does not assume that the perturbations have a small amplitude, thereby accounting for the total curvature perturbation. However, mapping non-linear effects is not obvious in the $\Delta N$ formalism, as we shall see shortly. When non-Gaussianity is present in the tail of the distribution, it does not reflect in the variance of the distribution (equivalently, the power spectrum). Instead, one must rely on the three-point and four-point correlation functions, or equivalently, the bispectrum and trispectrum respectively \cite{Bartolo:2004if}. In this work, we restrict ourselves to the study of the non-perturbative power spectrum and reserve the study of higher correlation functions via the $\Delta N$ formalism for follow-up work. We begin by reviewing the algorithm to compute the power spectrum via stochastic $\Delta N$ formalism, first studied in \cite{Fujita:2013cna, Fujita:2014tja}. As $N$ is the time variable in our system of equations, the recorded first passage time is understood as the e-fold corresponding to the case when the background field reaches a specific value for the first time (in this case,  the end of inflation), i.e.
\begin{align} \label{FPT}
     \phi_{\rm FPT} (N) &= \phi_{f}~.
\end{align}
\noindent We know from \cref{Stochastic_delta_N} that it should be possible to get the two-point statistics of the perturbations by tracking the time evolution of these independent constant density hypersurfaces. Fourier transformation of \cref{eq:R_N_relation} gives,
\begin{align}\label{eq:R_delta_N}
    \int_{0}^{\infty} d{\rm k} {\cal R}_{\rm \textbf{k}} (N) e^{i {\rm \textbf{k}} \cdot x} &= \int_{0}^{\infty} d{\rm k}(N_{\rm \textbf{k}} - \Bar{N}_{\rm \textbf{k}}) e^{i\textbf{k} \cdot x}=  \int_{0}^{\infty} d{\rm k} \Delta N_{\rm \textbf{k}}  e^{i\textbf{k} \cdot x}~, \\
    {\cal R}_{\rm \textbf{k}} (N) &= \Delta N_{\rm \textbf{k}} ~,\label{delta_N}
\end{align}

\noindent where $\Bar{N}$ is the average over the ensemble of all the stochastic realizations. The noise is zero-centred and Gaussian so, we expect $\Bar{N}$ to coincide with the classical trajectory. After running a finite number of stochastic realizations for a given model, we obtain the variance of $N$ from the ensemble of the realizations of $N_{\rm f} = N(\phi_{\rm FPT})$ by

\begin{align}
    \langle \Delta N_{\rm \textbf{k}}^2 \rangle &= \langle (N_{{\rm f},{\rm \textbf{k}}}-\Bar{N}_{{\rm f},{\rm \textbf{k}}}^2 \rangle~,  \nonumber \\
    &= \langle N_{{\rm f},{\rm \textbf{k}}}^2 - 2\Bar{N}_{{\rm f},{\rm \textbf{k}}} N_{{\rm f},{\rm \textbf{k}}} + \Bar{N}_{{\rm f},{\rm \textbf{k}}}^2 \rangle~, \nonumber \\
    &= \langle N_{{\rm f},{\rm \textbf{k}}}^2 - \Bar{N}_{{\rm f},{\rm \textbf{k}}}^2 \rangle~,
\end{align}


\noindent equating this quantity to the dimensionless primordial curvature power spectrum via
\begin{align}\label{power_spec_k}
    \langle \Delta N_{\rm \textbf{k}}^2 \rangle &= \int_{{\rm k}_i}^{{\rm k}_{\rm f}} \frac{d{\rm k}}{{\rm k}} {\mathcal P}_{\cal R}({\rm k}, \bar{N})~,
\end{align}
\noindent where ${\rm k}_i$ is the largest mode whose evolution starts at the beginning of the simulation (which is the first mode to cross the coarse-graining scale) and ${\rm k}_{\rm f}$ is the last mode that crosses the coarse-graining scale before the inflation ends. Therefore, all the modes in the interval ${\rm k}_i < {\rm k} <{\rm k}_{\rm f}$ contribute to the integral of \cref{power_spec_k} at the instant they cross the coarse-graining scale. As discussed in the \cref{testing_section}, it is relatively straightforward to obtain the variance of the probability distribution when working in real space. However, it is less obvious when we go to the Fourier space. This becomes evident when we differentiate \cref{power_spec_k} with respect to the mean e-folds $\bar{N}$ (or N-derivative) using the Leibniz integral rule. After ignoring the SR corrections i.e.  ${\rm k} \frac{d}{d {\rm k}}= \frac{1}{1 - \epsilon} \frac{d}{dN} \approx \frac{d}{dN}$, we get

%
\begin{align}\label{power_spec_delta_N}
 \left.  \frac{d \langle \Delta N_{\rm \textbf{k}}^2 \rangle}{d \Bar{N}} \right|_{\bar{N} = \ln({\rm k}_{\rm{f}}/{\rm k})} &= \left. {\mathcal P}_{\cal R}({\rm k}, \bar{N}) \right|_{\bar{N} = \ln({\rm k}_{\rm{f}}/{\rm k})} +   \int_{{\rm k}_i}^{{\rm k}_{\rm f}} \left( \frac{\partial  {\mathcal P}_{\cal R}({\rm k}, \bar{N}) }{\partial \bar{N}} \right) \frac{d{\rm k}}{{\rm k}}
\end{align}
%
 \noindent where the subscript denotes that the quantities are computed at the spatial hypersurface ${\rm k}_{\rm f} (N) = \sigma a (N) H (N)$. In the SR scenario, ${\mathcal P}_{\cal R}$ is nearly scale-invariant and freezes out until it reaches the coarse-graining scale. In this case, the second term on the right-hand side of \cref{power_spec_delta_N} vanishes, and the N-derivative of the variance coincides with the full power spectrum or the first term on the right-hand side of \cref{power_spec_delta_N}.  On the other hand, when the potential has an USR regime, the curvature perturbations are no longer conserved at super-Hubble scales and continue to grow proportionally to how long they are influenced by the USR phase \cite{Cheng:2018qof}. A significant consequence of this is that the second term in the right-hand side of \cref{power_spec_delta_N} is nonzero. The magnitude of this term not only depends on the rate of change of the power spectrum but also on the choice of the coarse-graining scale ${\rm k}_{cg}$. Hence,  It is not trivial to numerically compute this term as the integral itself requires computing the evolution of the full power spectrum evolution beforehand.

 \noindent This additional term in the non-perturbative approach exists because of reverse-mapping a time-varying Fourier space quantity (power spectrum) to the time derivative of a real space quantity (variance) and, it is crucial to retain the scale-dependence of ${\mathcal P}_{\cal R}$ \cite{Ando:2020fjm,Prokopec:2019srf, Cruces:2024pni}. In this section, we adopt the so-called time-independent way by omitting this term and referring to the corresponding power spectrum as the Super-Hubble Time-Independent Non-perturbative Spectrum (STINS). Notably, in the SR model, ${\mathcal P}_{\cal R} \approx {\mathcal P}^{\rm lin}_{\cal R}$. Furthermore, it is theoretically expected that any deviation in the stochastic power spectrum \cref{power_spec_delta_N} from the one obtained via quantum field theory \cref{eq:mukhanov_sasaki} should be beyond linear order in the SR parameters even in the presence of an USR regime \cite{Cruces:2018cvq}.  

%
%

\subsection{Quadratic potential}\label{non_pert_quadratic_potential}
A preliminary potential to begin with is a quadratic potential with just one free parameter
\begin{align}\label{quadratic_potential}
    V(\phi) &= \frac{1}{2}m^2 \phi^2~,
\end{align}
that features a standard SR regime. The field mass $m$ is chosen to be $ \mathcal{O}( 10^{-6}) \ M_{\rm p}$ to match the amplitude of the power spectrum at pivot scales with CMB observations \cite{Planck:2018jri}, where $M_{\rm p}$ is the Planck mass.

\begin{figure}[htb]
    \centering
    \includegraphics[width=\linewidth]{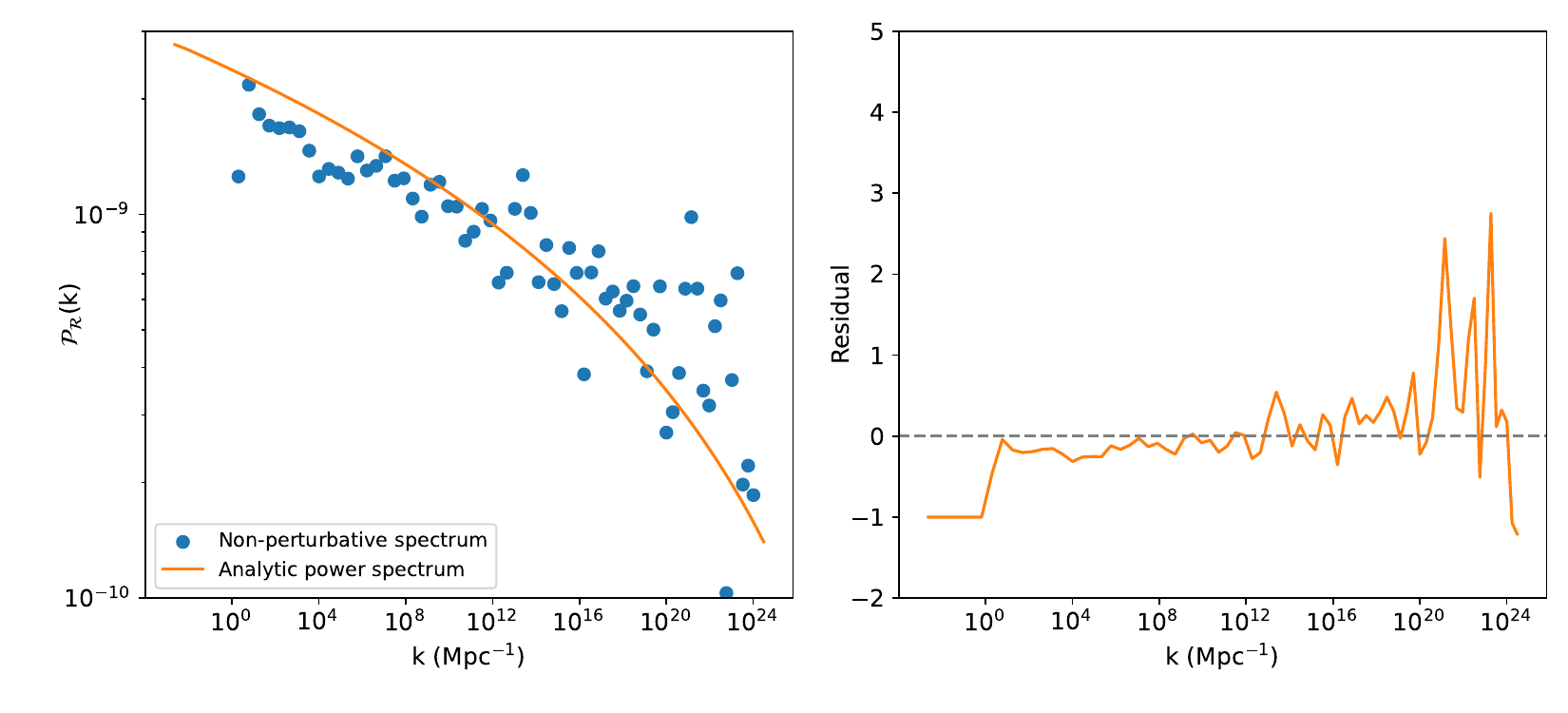}
    \caption{\textbf{Left Panel:} Dimensionless curvature power spectrum as a function of the comoving wavenumber $\rm k$ for the quadratic potential \cref{quadratic_potential}, computed from the slow-roll approximation (orange solid line) versus the non-perturbative power spectrum computed using \cref{power_spec_delta_N} (blue dots). \textbf{Right Panel:} The relative residual of the left panel.}
    \label{fig:power_spectrum_m2phi2}
\end{figure}

\noindent The curvature power spectrum is plotted in the left panel of \cref{fig:power_spectrum_m2phi2}. The analytic power spectrum at horizon exit is obtained from the fairly accurate slow-roll approximation in the attractor case with \cite{Baumann:2009ds, Senatore:2016aui}
\begin{align}\label{power_spect_analytic}
 \left. \mathcal{P_R} \right|_{SR} &= \frac{1}{24 \pi^2} \left. \left( \frac{V}{\epsilon} \right) \right |_{{\rm k}=aH} ~.  
\end{align}
\noindent The dotted blue point is our numerical result, which corresponds to the numerical implementation of the non-perturbative definition of the power spectrum \cref{power_spec_delta_N}. Each dot denotes a variation in $N$ from its mean on a specific point on the field-spacing grid $\left\{ \phi (\bar{N}_{\rm i}), ..., \phi (\bar{N_{\rm f}}) \right\}$ The dip at the first few e-folds is because the largest mode that exits the horizon first is supposed to be deep enough inside the physical horizon, i.e. $ \left. \frac{{\rm k}_{\rm min}}{aH}\right |_{t=t_{\rm i}}\gg1 $. Therefore, it takes some moments for the mode to cross the coarse-graining scale and kick the background. The spacing is carefully constructed such that a finite number of modes cross ${\rm k}_{\rm cg}$ within the interval $\left[\phi_{\rm j} (\bar{N}) - \phi_{{\rm j}-1} (\bar{N})\right]$. This grid is constructed on a classical trajectory (hence the bar on the e-folding variable) otherwise the field itself would be a stochastic variable. As discussed in \cref{Stochastic_intro}, this variation in the expansion rate of separate universe patches at a specific $\phi(\bar{N}_{\rm j})$ slicing is proportional to the consolidated outcome of the kicks generated up to that slicing since the initial hypersurface i.e.
\begin{align}\label{consolidated_noise}
   \left. \langle \Delta N_{\rm k}^2 \rangle \right|_{N = \bar{N}_{\rm j}} &\propto \int_{\phi ({\bar{N_{\rm i}}})}^{\phi (\bar{N_{\rm j}})} \xi (\bar{N}) d \bar{N}~,
\end{align}
Note that it is not a linear relationship because of the non-linear nature of the system. Moreover, the right-hand side in \cref{consolidated_noise} represents both field and momentum noises. Now, the residual plot in the right panel of the \cref{fig:power_spectrum_m2phi2} becomes intuitive. It closely follows zero in the beginning, signalling an excellent agreement between the two approaches. After about 40 e-folds, the non-perturbative spectrum shows minor deviation although comfortably below one until $\bar{N} \sim 50$ after which the discrepancy increases gradually. This increasing disagreement is attributed to the fact that at later times more noise has been assimilated into the background evolution and hence higher uncertainty compared to earlier $\phi(\bar{N})$ slices, as evident from \cref{consolidated_noise}. This causes more sudden fluctuations in the progress of the last few e-folds before the inflation ends. Still, the residual at its maximum does not go beyond a factor of three, which shows the robustness of the non-perturbative method. Note also that close to the end of inflation, the condition $\epsilon \ll 1$ is not strictly followed, and the analytic power spectrum \cref{power_spect_analytic} gets corrections beyond $\mathcal{O}(\epsilon)$ \cite{Stewart:1993bc}. More information about the accuracy of the results can be found in \cref{accuracy check_appendix}.

\noindent We add a last interesting note before ending this section: The variance of the noise is directly proportional to the energy scale of inflation, as seen from \cref{field_noise_variance_SW,momentum_noise_variance_SW}. Hence, choosing a higher mass of inflaton would elevate the potential energy \cref{quadratic_potential} thus, the variance. A larger variance would mean that high amplitude kicks will be more probable whose effect will be more clearly noticeable by the numerical integrator compared to the tinier kicks. In such a case, it would be computationally cheaper to resolve the non-perturbative power spectrum although the amplitude will not be in line with the CMB observations. On the other hand, if the variance is too large, it can cause the stochastic integrator to get stuck forever without the field reaching the end of inflation.

\subsection{Realistic potential: USR feature}\label{sec:Realistic_Potential_power_spectrum}

We revisit the potential discussed in \cref{realistic_potential_section} with the same choice of model parameters. The pivot scale is adjusted at about 55 e-folds before the end of inflation to match the power spectrum with the Planck scales \cite{Planck:2018jri}. As seen from the Langevin equation \cref{pi_eq} and \cref{noise_evolution_plots}, the deterministic drift term primarily drives the field push along the SR trajectory. Although the diffusion term is nonzero, it is much suppressed and thus for an arbitrary realization, the stochastic velocity trajectory does not deviate much from the classical trajectory. Hence, the field velocity at the onset of the USR is crucial to determine the probability that the field escapes the USR phase \cite{Pattison:2021oen}.


\noindent

\begin{figure}[htb]
    \centering
    \includegraphics[width=\linewidth]{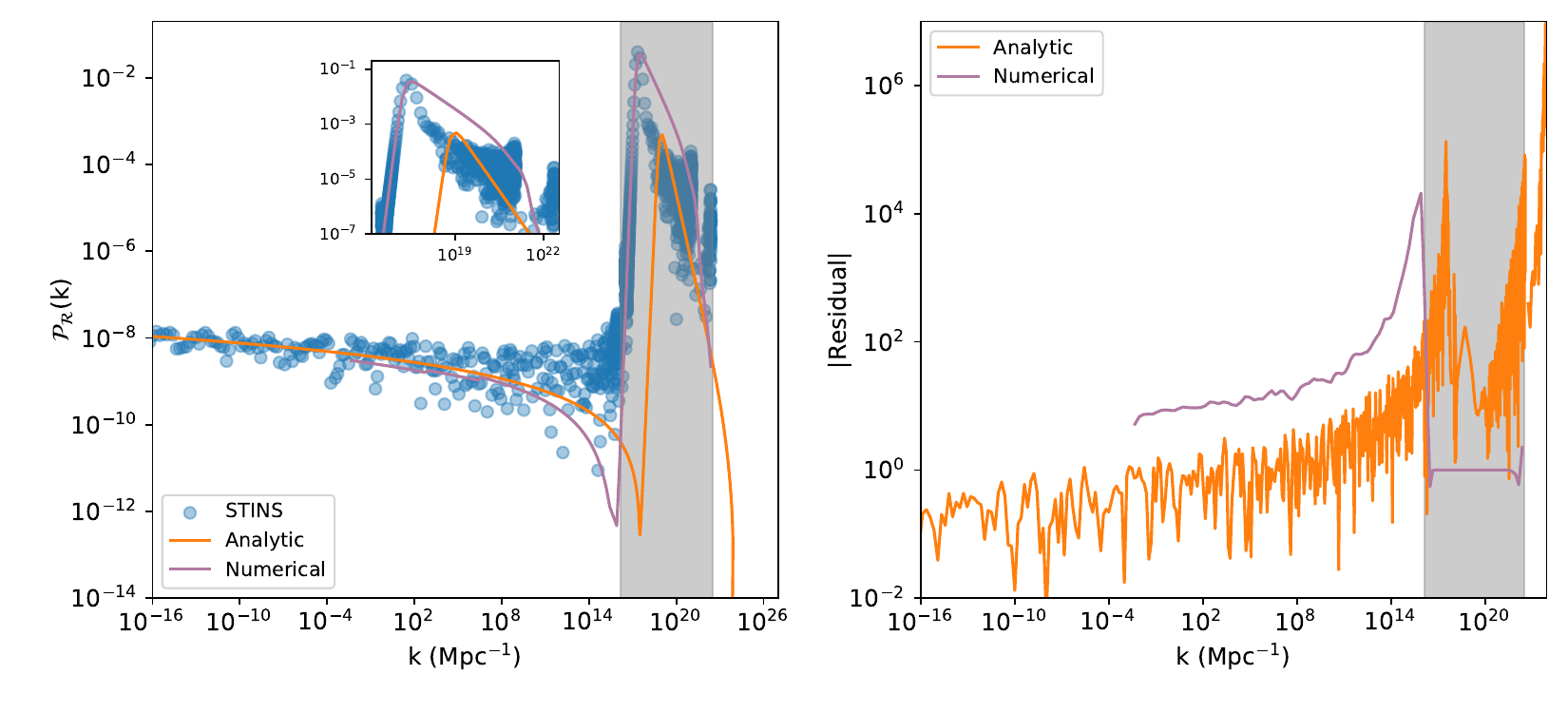}
    \caption{\textbf{Left Panel:} Dimensionless curvature power spectrum as a function of the comoving wavenumber ($\mathrm{k}$) for the potential featuring the USR regime (grey band) \cref{eq:realistic_potential}. The STINS computed using \cref{power_spec_delta_N} (blue dots) is shown along with the analytic formula (blue solid line) and the numerical solution to the Mukhanov-Sasaki equation. \textbf{Right Panel:} Absolute residual of the non-perturbative power spectrum in the left panel, relative to the analytic and numerical counterparts.}
    \label{fig:power_spectrum_realistic}
\end{figure}

\noindent We present our results for the STINS in \cref{fig:power_spectrum_realistic}. We also show the analytic power spectrum \cref{power_spect_analytic} and the numerical power spectrum computed using the Mukhanov-Sasaki equation \cref{eq:mukhanov_sasaki}, with the latter calculated from the moment the pivot scale (${\rm k}_* = 0.05 \ {\rm Mpc^{-1}}$) exits the horizon. In the right panel of the same figure, we plot the absolute residuals as functions of the comoving wavemodes. The STINS agrees well in the regime, which is far from non-attractor, as indicated by the residual barely exceeding an order of one magnitude.

\noindent The slow-roll-approximated power spectrum fails to capture the growth of perturbations in the USR regime \cite{Ballesteros:2017fsr,Motohashi:2017kbs}, which is why we rely on the solution of the Mukhanov-Sasaki equation in this regime. As the field approaches the USR regime, the residuals rise gradually reaching up to four orders of magnitude at the onset of USR, because the dip in the power spectrum just before the transition ${\rm SR} \rightarrow {\rm USR}$ is not as steep in the non-perturbative case compared to the numerical case. During the USR regime, the non-perturbative power spectrum tracks the rising slope of the numerical power spectrum with high accuracy. The peak amplitude of the two power spectra is of the order $\mathcal{O}(10^{-1})$. After the peak, the falling slope in the non-perturbative power spectrum narrows for some scales but eventually rises to catch up with the numerical case. The residual spikes again, albeit less pronounced, at the transition ${\rm USR} \rightarrow {\rm SR}$, shortly before inflation ends.

\noindent These discrepant regimes can be understood by observing how the absence of the second term of \cref{power_spec_delta_N} in our analysis affects the STINS relative to the (numerical) full power spectrum. The scales smaller than the dip grow with a steep positive slope i.e. $\frac{\partial {\mathcal P}_{\cal R} }{\partial \bar{N}} > 0$ at the onset of USR  (see \cite{Briaud:2025hra} for further discussion). This leads to the overestimation of the amplitude in the STINS compared to the numerical power spectrum, which is visible in the vicinity of ${\rm k} \sim 10^{16} \ {\rm Mpc^{-1}}$. Conversely, for the scales crossing the horizon just after the peak, ${\rm k} \gtrsim 10^{18} \ {\rm Mpc^{-1}}$, the power spectrum has a negative slope, although less steep than at the USR entry. Hence, $\frac{\partial {\mathcal P}_{\cal R} }{\partial \bar{N}} < 0$ implies that the STINS would underestimate the amplitude in this region as reflected in the grey band of \cref{fig:power_spectrum_realistic}. Thus, the non-vanishing time-dependent term in \cref{power_spec_delta_N} becomes very crucial, particularly when the slope of the power spectrum is steep near the local extrema. The residual plot in \cref{fig:power_spectrum_realistic} quantifies the scale-wise significance of the time-dependence of the curvature power spectrum. These features are fairly stable against the numerical accuracy settings, as verified in \cref{accuracy check_appendix}. Since the abundance of PBHs is exponentially sensitive to the enhancement in the power spectrum, the importance of the time-dependent term cannot be neglected if one needs to rely on a non-perturbative approach for precise PBH statistics. Addressing this issue will be a key focus of our future work.

\section{Conclusion}\label{conclusion}
A concrete perception of the correlation functions of primordial curvature perturbations is crucial for properly understanding PBHs, whose formation and population are sensitive to them. In this paper, we harness the capability of stochastic inflation and the $\Delta N$ formalism to address this issue non-perturbatively. For this purpose, we have utilized a numerical algorithm to simulate numerous stochastic realizations of certain inflationary models, built upon the statistics of the first-passage time of the end of inflation e-folds.

To validate our code, we have chosen two classes of models and obtained the probability distribution of the stochastic variable $N (\phi_{\rm f})$: first, a cubic-tilted potential for which we successfully verified the results of Ref. \cite{Ezquiaga:2019ftu, Figueroa:2021zah} for the case where the inflation duration is short i.e. $(\alpha=0.04)$, as evident from the left panel of \cref{fig:PDF_toy_model}. In the other case $(\alpha=0.02)$, where inflation lasts longer, we notice a marginally undervalued variance of $N$ compared to the mentioned known results. We argue that this should be a result of the backreaction effect that occurs when the noise is computed on a stochastic background. We also provide a comparison of the full noise treatment with the simplified treatment where the field variance is given by \cref{simplified_field_noise_variance_SW} and momentum noise is turned off. Secondly, we choose a physically motivated potential \cite{Ballesteros:2024pwn} with an ultra-slow roll (USR) region close to the end of inflation.  In this case, too, we find an excellent agreement of the exponential tail obtained from our code up to $\Delta N \approx 2.5$.

In \cref{non_pert_algo}, we have presented our primary results, translating the algorithm for constructing a non-perturbative power spectrum to a numerical method and applying it to study two different inflationary models. For the quadratic potential case, we confirm the scale-invariant curvature power spectrum via the non-perturbative approach, showing percent-level agreement with the analytic power spectrum. Additionally, for the USR model, we have obtained the peak in the power spectrum at small scales, which matches the magnitude of the peak obtained in the perturbative power spectrum. This is expected because, although the effects of quantum diffusion lead to an enhanced probability of very rare realizations, the power spectrum only depends on the variance of the distribution. We also find that the Super-Hubble Time-Independent Non-perturbative Spectrum (STINS) effectively captures the key features of the perturbations in the USR case but shows limitations in certain regimes. This happens because the perturbations keep growing even after horizon crossing, as long as the field is in the non-attractor phase. Consequently, the time-independent assumption is no longer valid and $\frac{\partial {\mathcal P}_{\cal R} }{\partial \bar{N}} \ne 0$. Owing to this, a discrepancy between STINS and the numerical power spectrum is observed in the regions where the spectrum changes rapidly over a short period. By analyzing residuals between STINS and the perturbative power spectrum, we identify at what scales this approximation is less accurate, providing insight into the regimes where accounting for the time-dependence of the power spectrum is essential. 

To conclude, multiple extensions can be built on our proposed algorithm, which is already very generic to single-field models. Firstly, we plan to obtain the full non-perturbative spectrum by also incorporating its time-dependent component in \cref{power_spec_delta_N}. Additionally, the non-perturbative bispectrum and trispectrum of the curvature perturbations can be numerically computed through the higher moments of the statistical variable $\Delta N$ \cite{Vennin:2015hra,Kalaja:2019uju}. However, as pointed out in our work, the time dependence of the USR models adds complexity to such computations. Another key task is quantifying the backreaction effects of the stochastic noise discussed in \cref{testing_section}, a topic we plan to explore in the future.

\section*{Acknowledgements}

The author is grateful to Julien Lesgourgues for several insightful discussions and for proofreading. The author also thanks Christian Fidler for discussions about stochastic processes and Joseph Jackson for discussions about the code. DS is thankful to Guillermo Ballesteros for clarifications regarding the USR model. DS also thanks Diego Cruces, Danilo Artigas and the anonymous referee for helpful comments and feedback on the stochastic $\Delta N$ formalism. Computations were performed with computing resources granted by RWTH Aachen University under project rwth1672. DS is supported by the German Academic Exchange Service (DAAD) grant.

\appendix

\section{The Mukhanov-Sasaki equation}\label{SW_inaccuracy}

In order to derive reach \cref{KG_eq_pert} we start from the Klein-Gordon equation \cite{Pattison:2019hef}:

\begin{align}\label{KG_eq_BG}
    \left( \frac{\partial}{\partial t^2} + \frac{\partial}{\partial {x^i}^2}\right) \phi({\rm{x}}^i,t) + 3 H(t) \frac{\partial \phi({\rm{x}}^i,t)}{\partial t} &= \frac{\partial V(\phi({\rm{x}}^i,t))}{\partial \phi}~,
\end{align}

 \noindent in the presence of perturbations, the inflaton becomes a localized variable, and it can be decomposed into a homogeneous background variable plus the perturbed variable at linear order

\begin{align}\label{delta_phi}
   \phi({\rm{x}}^i,t) &= \bar{\phi}(t) + \delta \phi({\rm{x}}^i,t) ~,
\end{align}

\noindent we drop the variable dependency inside the braces for convenience. Upon substituting \eqref{delta_phi} in \eqref{KG_eq_BG}, one obtains the linearized Klein-Gordon equation in Fourier space \cite{Pattison:2019hef, Gordon:2000hv} 

\begin{align}\label{KG_eq_pert}
    \Ddot{\delta \phi_{\rm \textbf{k}}} + 3 H \Dot{\delta \phi_{\rm \textbf{k}}} + \left(\frac{{\rm k}^2}{a^2} + \frac{\partial^2 V}{\partial \phi_{\rm \textbf{k}}^2} \right) \delta \phi_{\rm \textbf{k}} &= - 2 \frac{\partial V}{\partial \phi_{\rm \textbf{k}}} A +\Dot \phi_{\rm k} \left(\Dot{A} + 3 \Dot{\psi} + \frac{\rm k^2}{a^2}(a^2 \Dot{E} - aB) \right)~,
\end{align}

\noindent the terms in the right-hand side of \eqref{KG_eq_pert} represent the metric perturbations of a generalized perturbed line element

\begin{align}
    ds^2 &= (1+A)dt^2 + 2a \frac{\partial B}{\partial x^i} dx^i dt + a^2 dx^i dx^j \left((1 - 2 \psi) \delta_{ij} + 2 \frac{\partial^2 E}{\partial x^i \  \partial x^j } \right)~,
\end{align}

\noindent neglecting the lapse function A signifies one is working in a gauge with the time coordinate fixed. If the number of e-folds is the time coordinate in this case, then the choice of gauge is referred to as uniform-N gauge. 

\noindent The behaviour of metric perturbations can be obtained by solving for the energy and momentum constraint equations of the Einstein field equations

\begin{align}\label{energy_constraint}
3H (\Dot{\psi} + H A) +   \frac{\rm k^2}{a^2} \left(\psi - H(aB - a^2 \Dot{E}) \right) + 4 \pi G \left( \frac{\partial V}{\partial \phi_{\rm \textbf{k}}} \delta \phi_{\rm \textbf{k}} + (\delta \Dot{\phi_{\rm \textbf{k}}} - \Dot{\phi}A) \Dot{\phi_{\rm \textbf{k}}}\right) &= 0~,
\end{align}

\begin{align}\label{momentum_constraint}
\Dot{\psi} + H A + 4 \pi G \Dot{\phi_{\rm \textbf{k}}} \delta \phi_{\rm \textbf{k}} &= 0~.
\end{align}

\noindent We work in the spatially flat gauge, where the spatial metric perturbation is switched off, i.e. $\psi= 0$. In this gauge, the field perturbations coincide with the Mukhanov-Sasaki variable, defined as

\begin{align}
    Q_{\rm \textbf{k}} &= \delta \phi_{\rm \textbf{k}} + \frac{\Dot{\phi_{\rm \textbf{k}}} \psi}{H}~,
\end{align}

\noindent and the constraint equations \eqref{energy_constraint}, \eqref{momentum_constraint} become

\begin{align}
    3H^2 A +   \frac{{\rm k}^2}{a^2} \left( - H(aB - a^2 \Dot{E}) \right) + 4 \pi G \left( \frac{\partial V}{\partial \phi_{\rm \textbf{k}}} \delta \phi_{\rm \textbf{k}} + (\delta \Dot{\phi_{\rm \textbf{k}}} - \Dot{\phi}A) \Dot{\phi_{\rm \textbf{k}}}\right) &= 0~, \label{energy_constraint_SFG} \\
        H A + 4 \pi G \Dot{\phi_{\rm \textbf{k}}} \delta \phi_{\rm \textbf{k}} &= 0~.\label{momentum_constraint_SFG}
\end{align}


\noindent Using \eqref{energy_constraint_SFG} and \eqref{momentum_constraint_SFG}, we can get rid of the metric perturbations on the right-hand side of \eqref{KG_eq_pert}, such that 

\begin{align}\label{eq:mukhanov_sasaki}
    \Ddot{\delta \phi_{\rm \textbf{k}}} + 3H \dot{\delta \phi_{\rm \textbf{k}}} + \left[  \frac{{\rm k}^2}{a^2} + \frac{\partial^2 V}{\partial \phi_{\rm \textbf{k}}^2} - \frac{8 \pi G}{a^3}\frac{d}{dt} \left( \frac{a^3 \Dot{\phi^2_{\rm \textbf{k}}}}{H}\right)\right] \delta \phi_{\rm \textbf{k}}&= 0~,
\end{align}

\noindent which is also popularly called the Mukhanov-Sasaki equation. Finally, we perform a change of variable $dt \rightarrow dN$

\begin{align}\label{SW_eqn_appendix}
\delta \phi_{\rm \textbf{k}}^{\prime \prime} +  (3 + H^{\prime}/H) \delta \phi^{\prime}_{\rm \textbf{k}} + \left({\rm k}^2 /(a H)^2+\bar{\phi^{\prime}_{\rm \textbf{k}}}^2\left(3+H^{\prime} / H\right)+2 \phi^{\prime}_{\rm \textbf{k}} V_{, \bar{\phi}} / H^2+V_{, \bar{\phi} \bar{\phi}} / H^2\right) \delta \phi_{\rm \textbf{k}} &=0~,
\end{align}

\noindent which shows a mismatch in the oscillation frequency of the \eqref{pert_KG_eq} used by \cite{Figueroa:2020jkf}.

\section{Noise treatment for the Probability distribution }\label{noise_evolution_plots}
\begin{figure}[htb]
    \centering
    \includegraphics[width=\linewidth]{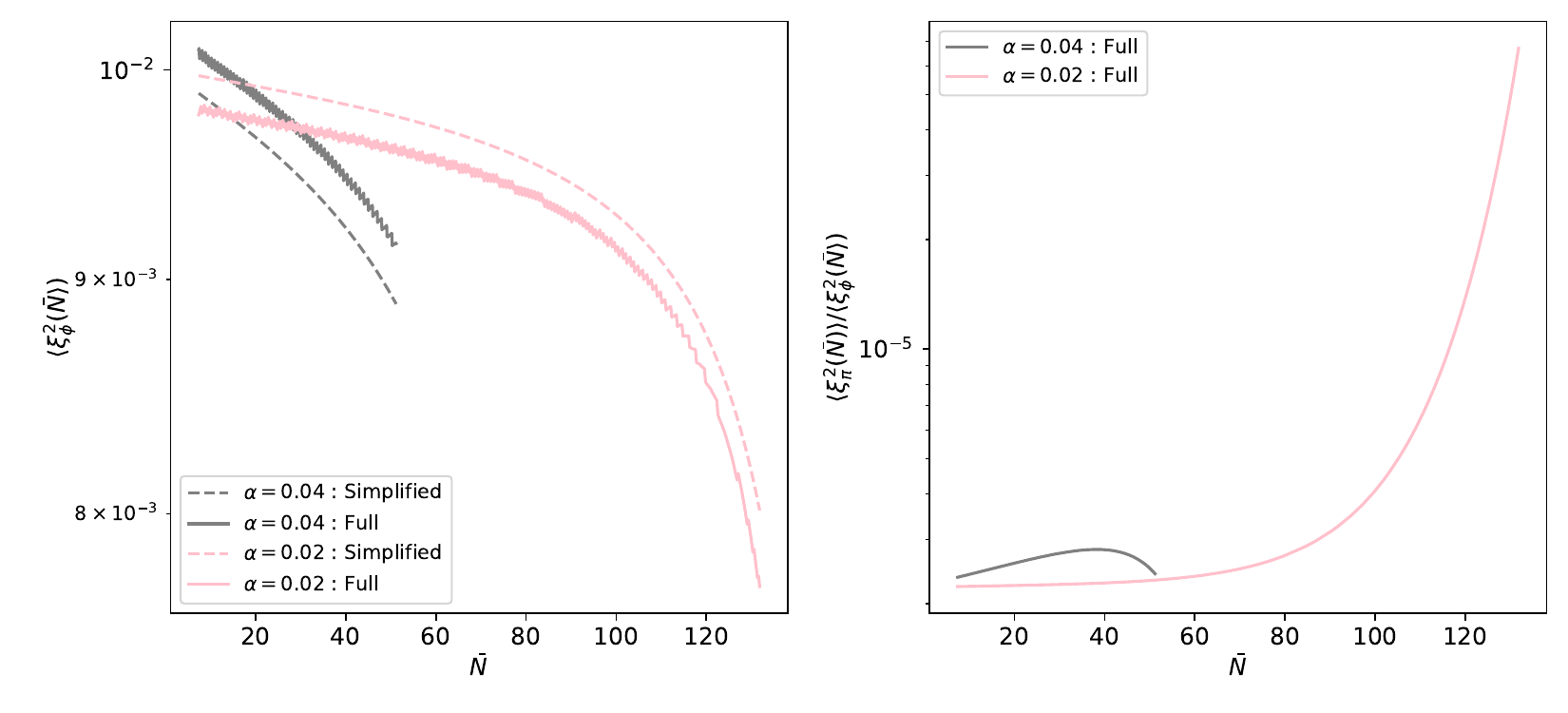}
    \caption{\textbf{Left Panel:} Field noise variance for the potential \cref{toy_potential} as a function of the mean e-folds, computed using full calculation (solid lines) \cref{field_noise_variance_SW} and compared with the simplified case \cref{simplified_field_noise_variance_SW} (dashed lined) for two different values of $\alpha$ parameter. \textbf{Right Panel:} Same as in the left panel but for the momentum noise variance saturated with the field noise variance, which is also zero for the simplified case.}    \label{fig:noise_toy_model}
\end{figure}

\begin{figure}[htb]
    \centering
    \includegraphics[width=\linewidth]{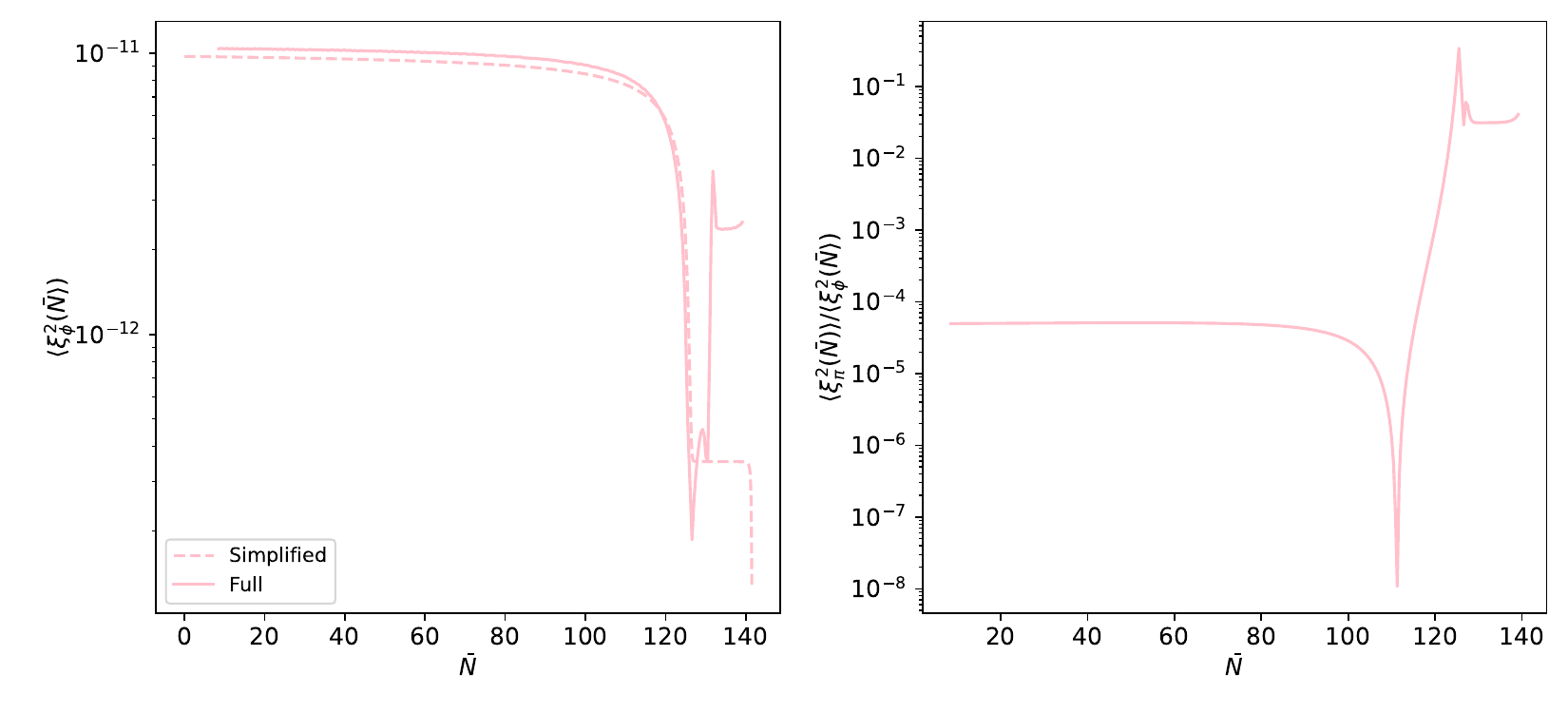}
    \caption{\textbf{Left Panel:} Field noise variance for the potential \cref{eq:realistic_potential} as a function of the mean e-folds, computed using full calculation (solid lines) \cref{field_noise_variance_SW} and compared with the simplified environment \cref{simplified_field_noise_variance_SW} (dashed lined). \textbf{Right Panel:} Same as in the left panel but for the momentum noise saturated with the field noise, which is also zero for the simplified case.}    \label{fig:noise_realistic_model}
\end{figure}

Here we show the noise correlators of the models studied in \cref{testing_section}. The noise for two cases of toy potential \cref{toy_potential} is plotted in the \cref{fig:noise_toy_model}. It is clear from the left panel that field noise is roughly of the same order of magnitude for both full treatment \cref{field_noise_variance_SW} and simplified treatment \cref{simplified_field_noise_variance_SW}. Moreover, the momentum noise variance is suppressed by five orders of magnitude as opposed to the field noise variance as shown in the right panel. For the case of the realistic potential \cref{eq:realistic_potential}, the noise correlator is shown in the \cref{fig:noise_realistic_model}. It is visible in the left panel of this figure that the simplified treatment fails to capture the non-trivial feature of the field noise in the USR regime. Furthermore, the momentum noise variance reaches almost at the same order in magnitude as the field noise variance in the USR regime, implying that it is incorrect to ignore the momentum noise in the USR regime.


%
\section{Numerical accuracy check for non-perturbative spectrum}\label{accuracy check_appendix}

\begin{figure}[htb]
    \centering
    \includegraphics[width=0.9\linewidth]{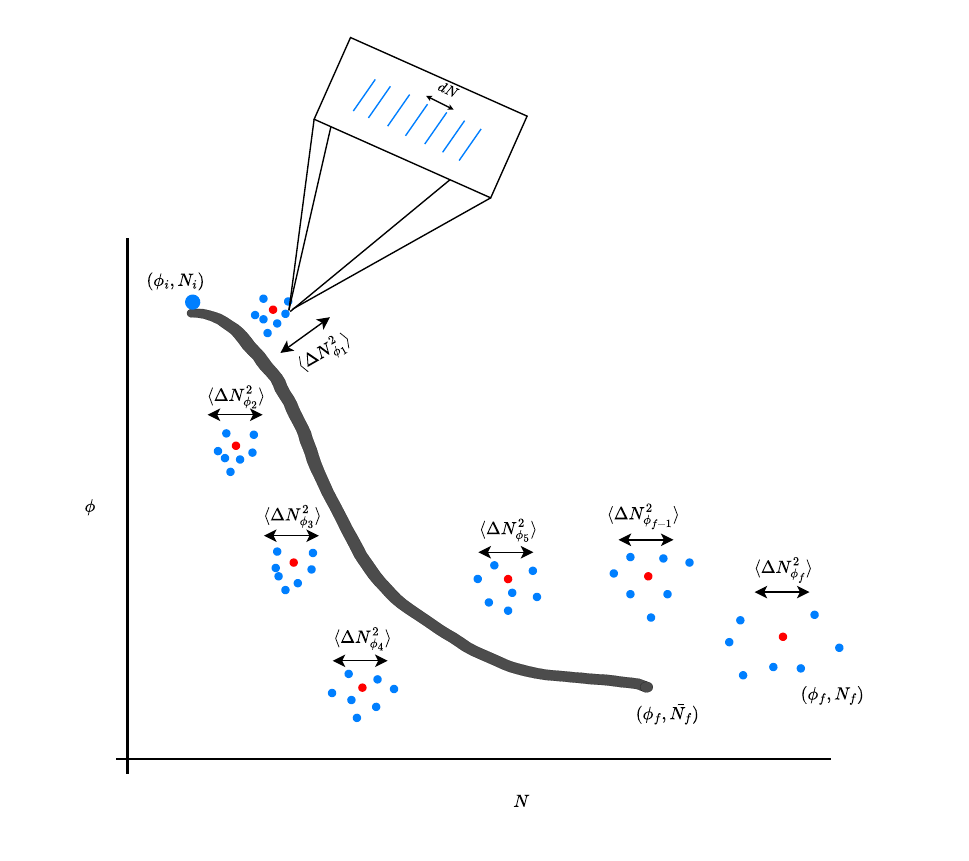}
    \caption{An 
    ensemble of magnified stochastic trajectories of the field showcasing the concept behind the non-perturbative power spectrum. Red dots are the mean e-folds $\bar{N_{\phi}}$ mapped to the field spacing and blue dots correspond to the sample realizations. The time grid for the realizations has a stepsize $dN$.}
    \label{fig:accuracy_check_diagram}
\end{figure}

The diagram in \cref{fig:accuracy_check_diagram} sums up the idea of numerically implementing the technique of \cref{non_pert_algo}. The field-spacing grid $\left\{ \phi (\bar{N}_{\rm i}), ... , \phi (\bar{N_{\rm f}}) \right\}$ stores the stochastic variable $N$ at each predefined point from initial hypersurface $\phi(\bar{N_{\rm i}})$ until the end of inflation hypersurface $\phi(\bar{N_{\rm f}})$ as the field rolls down the potential. The variances in $N$ at the individual $\phi$ slicing are shown as $\langle \Delta N^2_{\phi_{\rm j}} \rangle$. Since the noises have zero mean value, the mean e-folding at each stage coincides with the classical e-folding parameter at that stage i.e. $\langle N_{\phi_{\rm j}} \rangle = \bar{N}(\phi_{\rm j})$, which is shown as red dots in the diagram. It does not mean the statistical data will record such a classical realization, it is just for representation purposes. The field keeps stopping further from the mean value at later times due to integrated noise effects explained in \cref{non_pert_quadratic_potential}, as is evident from the dots being shown more distant near the end of inflation. There is also a magnified picture of field evolution in an arbitrary realization, where the parameter $dN$ controls how far the field has to jump to reach the instantly successive spatial hypersurface. Hence, at the macro level, the number of realizations control how well we can estimate the spread of the $N$ distribution (or the variance) and at the micro level, the time step $dN$ controls how accurately the time evolution of each stochastic realization takes place. These two parameters (number of realizations $n_{\rm R}$ and the time step $dN$) are the primary parameters of interest when accurately constructing the non-perturbative power spectrum.

\subsection{Quadratic potential}
\begin{figure}[htb]
    \centering
    \includegraphics[width=\linewidth]{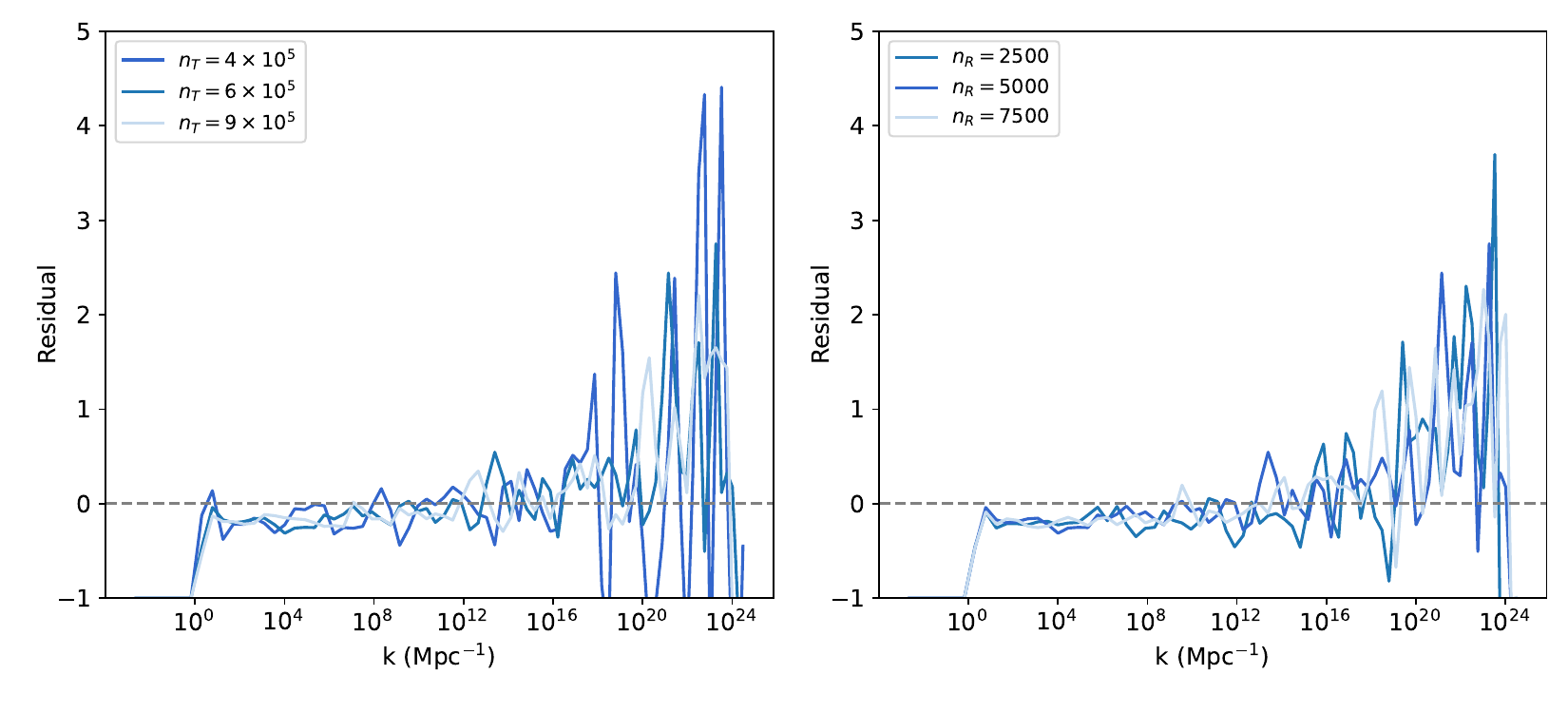}
    \caption{\textbf{Left Panel:}Residual plot for the non-perturbative power spectrum relative to the analytic counterpart for the quadratic potential \cref{quadratic_potential}, with $n_{\rm R} = 5000$ as a function of the comoving wavenumber ${\rm k}$. For convenience, we have defined $n_T = \frac{\bar{N}}{d N}$. \textbf{Right Panel:} Same as in left panel but for $n_T=6 \times 10^5$}
    \label{fig:m2phi2_accuracy_check}
\end{figure}
First, we check the sensitivity of the power spectrum in the case of the quadratic potential against varying the number of realizations $(n_{\rm R})$ and the time step $(dN)$. It is evident from the left panel of the \cref{fig:m2phi2_accuracy_check} that keeping the realizations constant, the residual has higher fluctuations for the larger time steps. For the smallest time step i.e. $dN = \frac{\bar{N}}{9 \times 10^5}$, the peak residual barely exceeds two, showcasing an excellent agreement of both the non-perturbative and analytic power spectra. On the other hand, we get a lower residual when we simulate more realizations with a constant time step. This is shown in the right panel of \cref{fig:m2phi2_accuracy_check}, where again, the residual stays within two with $n_{\rm R}=7500$. 
\subsection{Realistic potential}
\begin{figure}[H]
    \centering
    \includegraphics[width=\linewidth]{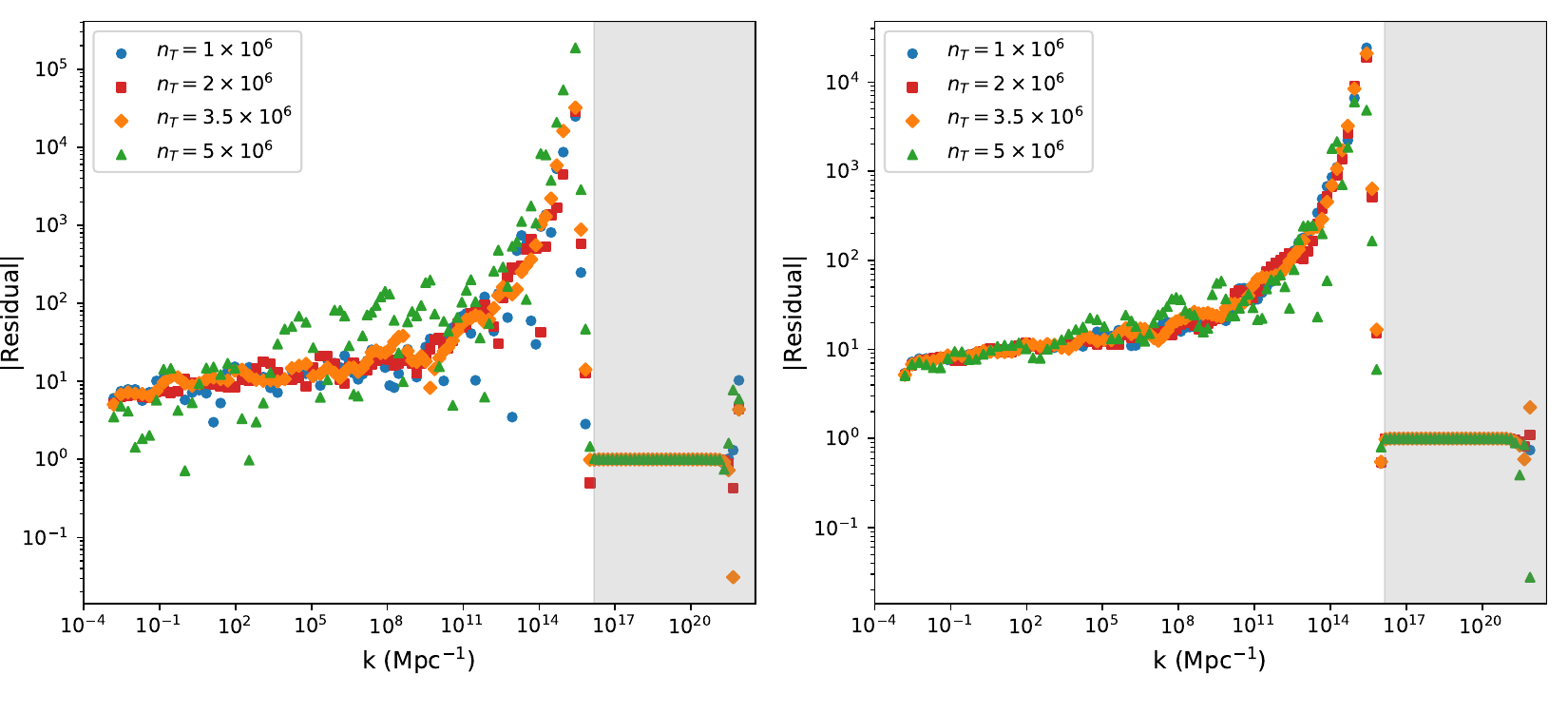}
    \caption{\textbf{Left Panel:}Residual plot for the STINS relative to the numerical power spectrum for the realistic potential \cref{eq:realistic_potential} with $n_{\rm R} = 750$ as a function of the comoving wavenumber ${\rm k}$. The grey band shows the USR regime. \textbf{Right Panel:} Same as in left panel but for $n_{\rm R}=5000$}
    \label{fig:realistic_accuracy_check}
\end{figure}
In the \cref{fig:realistic_accuracy_check}, we show the parameter sensitivity of the STINS for the USR model. Notably, the time step needs to be around an order smaller than the SR model to achieve good accuracy. Overall, the residual shows a similar trend with the comoving scales for all the cases. In the attractor regime, the residual does not change by more than one order after substantially increasing the number of realizations. Even though the residual at the transition improves notably, the discrepancy of the STINS with the numerical result is still high, suggesting that it is not an outcome of numerical inaccuracy. Moreover, increasing the number of realizations makes the results less sensitive towards $n_T$. Additionally, the field spacing grid has to be non-uniformly distributed in the USR case with much higher fineness needed in the USR regime than the attractor regime. The reason is simple; in the USR regime, the field value stays nearly constant but there is a steep slope in the noise variance and hence, we need many points to capture this trait. We tune the total field grid points between 2000 and 3500 to minimize the wiggles in the non-perturbative power spectrum for each parameter set in the \cref{fig:realistic_accuracy_check}. 

\bibliographystyle{utphys}
\bibliography{references}

\end{document}